\pdfoutput=1
\documentclass[11pt]{article}

\usepackage[margin=1in]{geometry}
\usepackage[T1]{fontenc}
\usepackage[utf8]{inputenc}
\usepackage{lmodern}
\usepackage{microtype}

\usepackage{amsmath,amssymb,amsfonts}
\usepackage{amsthm} 
\usepackage{booktabs}
\usepackage{graphicx}
\usepackage{longtable}

\usepackage{siunitx}
\usepackage[version=4]{mhchem}
\usepackage{algorithm}
\usepackage{algpseudocode}
\usepackage{float}

\usepackage[authoryear,round,sort&compress]{natbib}
\usepackage{csquotes}

\usepackage{xcolor}
\usepackage[colorlinks=true,linkcolor=black,citecolor=black,urlcolor=black]{hyperref}

\newcommand{\keywords}[1]{\par\noindent\textbf{Keywords: }#1\par}

\title{Right-to-Override for Critical Urban Control Systems: A Deliberative Audit Method for Buildings, Power, and Transport}

\author{Rashid Mushkani\\[4pt]
Universit\'e de Montr\'eal\\
Mila--Quebec AI Institute}

\date{} 

\begin{document}
\maketitle

\begin{abstract}
Automation now steers building HVAC, distribution grids, and traffic signals, yet residents rarely have authority to pause or redirect these systems when they harm inclusivity, safety, or accessibility. We formalize a \textit{Right-to-Override} (R2O)—defining override authorities, evidentiary thresholds, and domain-validated safe fallback states—and introduce a \textit{Deliberative Audit Method} (DAM) with playbooks for pre-deployment walkthroughs, shadow-mode trials, and post-incident review. We instantiate R2O/DAM in simulations of smart-grid load shedding, building HVAC under occupancy uncertainty, and multi-agent traffic signals. R2O reduces distributional harm with limited efficiency loss: load-shedding disparity in unserved energy drops from $5.61\times$ to $0.69\times$ with constant curtailment; an override eliminates two discomfort-hours for seniors at an energy cost of \SI{77}{\kilo\watt\hour}; and median pedestrian wait falls from \SI{90.4}{s} to \SI{55.9}{s} with a \SI{6.0}{s} increase in mean vehicle delay. We also contribute a policy standard, audit worksheets, and a ModelOps integration pattern to make urban automation contestable and reviewable.
\end{abstract}

\keywords{Urban AI governance; Right‑to‑Override; Deliberative audit; Contestability; Smart grid; Building automation; Adaptive traffic control; ModelOps; Cities}


\section{Introduction}

Cities increasingly operate through automated control. Building management systems optimize HVAC setpoints and ventilation, distribution utilities orchestrate demand response, and traffic signal controllers adaptively reallocate green time. Efficiency gains are real, yet risks are not borne evenly. When objectives and constraints are defined without attention to plural publics, systems can lengthen pedestrian waits near elder centers, exacerbate energy poverty, or complicate accessibility for wheelchair users \citep{Sovacool2016EnergyJustice,Selbst2019Sociotechnical} \citep{Mushkani2025ICML_LIVS}. Sector standards already define baseline safety and risk practice—e.g., IEC~61508 for functional safety and ISO~31000 for risk management \citep{IEC61508_2010,ISO31000_2018}—and AI governance frameworks stress transparency, documentation, and risk management (e.g., NIST AI RMF~1.0; ISO/IEC~42001 for AI management systems; and U.S. OMB M\textendash24\textendash10) \citep{NISTAI_RMF2023,NISTGenAIProfile2024,OMB_M2410_2024}. Recent law adds obligations for human oversight and remedies; for example, the EU AI Act requires effective human oversight for high‑risk systems and creates post‑market complaint mechanisms \citep{EUAIAct2024}. Yet across these instruments there is little operational guidance for \emph{when and how} residents and civic bodies may legitimately interrupt \emph{urban} automated control loops—and how such powers should be exercised so they remain safe, proportionate, and reviewable.

We reconceptualize human oversight from operator‑centric intervention to \emph{civic oversight}. Drawing on scholarship in accountability and participatory design \citep{Ananny2018Accounts,Lee2019WeBuildAI,Wong2023Toolkit,CostanzaChock2020} and safety work on interruptibility and shutdown \citep{Leike2017AISafetyGridworlds}, we introduce the \textbf{Right‑to‑Override (R2O)} and the \textbf{Deliberative Audit Method (DAM)}. R2O specifies roles, triggers, and validated fallback states that bound how control policies may be paused or rolled back when distributional harm or safety risks cross agreed thresholds. DAM inserts structured, participatory review into the lifecycle of urban control systems—before deployment, in shadow‑mode, and after incidents—and yields artifacts that support accountability over time \citep{Raji2020AuditGap}. Contestability is treated as a design requirement realized in code, configuration, and operating procedures, not solely in documentation \citep{kaminski2021right,Mitchell2019ModelCards,Gebru2021Datasheets,Renn2018RiskGov,Sculley2015TechDebt}.

We ground the need for R2O/DAM in domain obligations. In transportation, the MUTCD (11th~ed., 2023) and PROWAG set expectations for accessible pedestrian signaling and safe fallback behavior at crossings \citep{MUTCD11_2023,PROWAG2023}. In buildings, thermal comfort and ventilation constraints provide anchors for safe states when automation conflicts with occupant needs \citep{ASHRAE55_2020,ASHRAE6212019}. Public‑sector AI policy requires impact assessment, independent evaluation, ongoing monitoring, fail‑safes, and (where practicable) human alternatives \citep{OMB_M2410_2024,NISTGenAIProfile2024,EUAIAct2024}. R2O operationalizes these expectations at the level where harm is experienced—by residents—so that any override is both \emph{authorized} and \emph{technically safe}.

\paragraph{Contributions.} This paper contributes four advances:
\begin{enumerate}
  \item A formal R2O standard defining override authorities, admissible evidence, and domain‑specific safe fallback states (aligned with safety/risk norms and human‑oversight requirements) \citep{IEC61508_2010,ISO31000_2018,EUAIAct2024,OMB_M2410_2024}.
  \item The Deliberative Audit Method (DAM) and its artifacts (playbooks, checklists, review templates) to make civic participation feasible, repeatable, and auditable \citep{Raji2020AuditGap,Lee2019WeBuildAI,Wong2023Toolkit}.
  \item Empirical evaluation of R2O/DAM in three domains using simulations calibrated to realistic operating envelopes: smart‑grid load shedding, building HVAC under occupancy uncertainty, and multi‑agent traffic signal control—reporting distributional, accessibility, and efficiency outcomes.
  \item A ModelOps integration pattern that couples policy, documentation, and incident response to code and configuration so governance is “live” in deployment.
\end{enumerate}
Across cases we find that R2O reduces distributional harm with limited efficiency loss, and that DAM yields actionable oversight artifacts rather than post hoc narratives.

\section{Background}

Urban automation governance draws on civic participation and deliberative democracy, safety engineering and resilience, regulatory risk management, and algorithmic accountability. Participation frameworks in planning and civic technology outline ladders of influence and boundary‑crossing practices by which stakeholders shape consequential systems \citep{Arnstein1969,Fung2006,Turnhout2010Participation,CostanzaChock2020} \citep{MushkaniKoseki2025Habitat_StreetReview}. Practice‑focused methods aim to move participation earlier in the lifecycle so that problem framing and scoping remain contestable \citep{Wong2023Toolkit,SituateAI_CHI2024} \citep{mushkani2025wedesigngenerativeaifacilitatedcommunity,mushkani2025streetreviewparticipatoryaibased}.

Risk and governance standards have matured but leave operational interruptibility under‑specified. NIST’s AI Risk Management Framework emphasizes oversight, documentation, and incident response as pillars of trustworthy systems \citep{NISTAI_RMF2023}. ISO/IEC~42001:2023 establishes an AI management‑system standard that integrates governance and continuous improvement across the lifecycle \citep{ISO42001_2023}. Public law is shifting as well: the EU AI Act requires human oversight for high‑risk systems, including the capability to stop a system into a safe state \citep{EUAIAct2024}. In the United States, OMB Memorandum M‑24‑10 directs agencies to institute minimum safeguards for uses of AI that affect rights and safety, with clear internal authority to pause or restrict deployments \citep{OMB_M2410_2024}. Canada’s Directive on Automated Decision‑Making similarly institutionalizes pre‑deployment review and transparency obligations in the public sector \citep{CanadaAIA2020}. None of these instruments, however, fully specifies who outside the operating organization can exercise interruptibility, which evidentiary thresholds should trigger holds, or what constitutes a validated safe state in domain contexts.

Algorithmic accountability scholarship in HCI and STS documents gaps between high‑level principles and lived impacts \citep{Selbst2019Sociotechnical,Ananny2018Accounts}. Auditing has moved from research practice to regulatory expectation, with proposals detailing independence, criteria, and assurance for external audits \citep{TerzisVeale2024,LamEtAl2024AssuranceAudits}. Civil‑society and governmental guides surface pragmatic mechanisms—impact assessments, registers, participatory review—that help publics discover and contest harms \citep{kaminski2021right}. Incident infrastructures such as the AI Incident Database complement these ex ante tools by systematizing post‑incident learning \citep{PAI_AIIDB}. Yet audit regimes and incident repositories typically stop short of allocating enforceable powers to pause or revert operational control when risks materialize.

Safety engineering and resilience research foreground graceful degradation and the design of safe fallback states. Longstanding machinery‑safety standards define emergency‑stop and safe‑state principles (e.g., ISO~13850 and IEC~60204‑1) \citep{ISO13850_2015,IEC60204_2016}. Sector‑specific reliability rules offer analogous templates: North American grid standards prescribe emergency operations and automatic under‑frequency load shedding with predefined configurations and time‑bounded authority \citep{EOP011_4_2024,NERC_PRC006_5_2021}. Building domains encode comfort and ventilation bounds that can anchor fallbacks (ASHRAE~55 and 62.1) \citep{ASHRAE55_2020,ASHRAE6212019}. These practices richly specify safe‑state design but rarely connect those states to civic oversight or distributional triggers.

There is mounting evidence that single‑objective automation in urban control can degrade equity, accessibility, or safety when multi‑criteria goals and guardrails are absent. In traffic signal control, throughput‑maximizing strategies such as max‑pressure and reinforcement‑learning controllers improve network flow but can bias service without explicit constraints; equity‑aware variants are only recently emerging \citep{PressLight2019,Chu2019MA2C,Liu2023TDMP}. In power systems, empirical and causal studies show restoration and outage burdens falling disproportionately on lower‑income or otherwise vulnerable communities, underscoring the need for procedural protections that can interrupt automated control when distributional thresholds are breached \citep{Sovacool2016EnergyJustice,Ganz2023PNASNexus,WeiEtAl2023Unfairness}. 

Our work complements these strands by specifying \emph{procedural rights}, \emph{evidentiary thresholds}, and \emph{domain‑validated fallback states} that make automated urban control \emph{contestable during operation}. The proposed Right‑to‑Override clarifies who can pause or roll back control, on what evidence, and for how long; the Deliberative Audit Method supplies concrete practices for pre‑deployment walkthroughs, shadow‑mode trials, and post‑incident review that connect participatory governance to engineering artifacts \citep{Wong2023Toolkit,SituateAI_CHI2024,kaminski2021right}.

\section{Materials and Partner Organizations}

We study three settings where automated control intersects with public‑interest obligations: electricity demand response and load shedding, heating/ventilation in community buildings, and adaptive traffic signal control in corridors used by both pedestrians and transit. Materials comprise (1) synthetic datasets and configurable simulation environments, (2) reference implementations of baseline controllers and validated fallback policies, and (3) draft governance artifacts for DAM, including scenario walkthrough worksheets, escalation playbooks, and post‑incident review templates. 

\paragraph{Electricity materials.} The environment represents a distribution service area with feeders that serve heterogeneous customers, including protected services (elevators, small clinics). Inputs include synthetic load profiles, a stylized shortfall that requires curtailment, and group labels used only to compute distributional monitors during analysis. Controllers implement a cost‑minimizing dispatch and a rotational‑curtailment fallback with equity caps. Monitors report total energy not served, normalized harm by group, and exceedances of minimum service for protected loads.

\paragraph{Building materials.} The environment models a single community facility with envelope heat transfer, internal gains, and variable occupancy. Inputs include an occupancy schedule with evening programming for older adults, a weather file, and comfort/indoor‑air‑quality bounds derived from ASHRAE~55 and 62.1. Controllers implement a scheduler with night setbacks and a fallback that maintains comfort and ventilation whenever protected occupancies are present. Monitors track discomfort‑hours during occupied periods and whole‑building energy use.

\paragraph{Transport materials.} The environment is a $4\times4$ arterial grid with signalized intersections, marked crosswalks, and one transit corridor. Inputs include demand profiles for vehicles and pedestrians and a headway target for the bus line. Controllers implement a throughput‑oriented adaptive policy and a fallback that provides a pedestrian walk phase every cycle and transit signal priority along the corridor. Monitors include delay for vehicles, wait time for pedestrians near sensitive sites, and headway deviation for buses \citep{Mushkani2025JUM_MontrealStreets}.

\paragraph{Partner organizations and roles.} A steering group guided the selection of monitors, default thresholds, and fallback definitions. The group included municipal practitioners (traffic operations, public buildings, electricity planning) and leaders from community organizations representing older adults, people with disabilities, and tenants of subsidized housing. Engagements were framed as organizational consultations rather than research with individuals. Partners helped to: (a) identify protected services and priority locations that should anchor safe fallbacks; (b) refine the phrasing of distributional, accessibility, safety, and service‑quality monitors; (c) stress‑test escalation rules through tabletop exercises; and (d) review the clarity and completeness of audit worksheets and public notices. Partners did not provide operational logs, customer records, or other sensitive data, and they did not endorse any particular numeric threshold used in the simulations.

\paragraph{Engagement procedure.} Sessions followed DAM’s sequence: scoping and role definition, scenario walkthroughs using synthetic maps and time series, and tabletop drills that rehearsed escalation and notification pathways. Notes were taken in non‑attributable form and analyzed thematically to surface recurrent requirements and concerns. These themes were translated into monitors, candidate thresholds for sensitivity analysis, and plain‑language definitions of safe fallbacks for each domain. 

\paragraph{Scope and limitations.} The purpose of the engagements was to ensure that materials, monitors, and fallback designs reflect practical constraints and community priorities. Because settings are simulated and data are synthetic, results should be interpreted as estimates of relative effects under the proposed R2O gating rather than as forecasts for any particular city. 

\section{Right‑to‑Override: Standard and Formalization}

\subsection{Authority, scope, and time‑bounds}

R2O defines three classes of action, each with pre‑validated scope and time‑bounds. \emph{Level 1} is the operator stop: duty engineers may pause an algorithm or revert to a safe fallback when there is an immediate safety or integrity risk. The pause is time‑limited to four hours and must be logged and reviewed. \emph{Level 2} is the municipal pause: a designated municipal controller may pause a policy for up to seventy‑two hours when equity, accessibility, or service‑quality monitors exceed adopted thresholds. \emph{Level 3} is the civic board hold: a standing civic board with community representation may impose a hold for up to thirty days when impacts are material, persistent, or systemic. Level~\!3 actions include requirements for mitigation plans and public notice. These levels parallel safety cases in critical systems and distinguish urgent risk stops from deliberative holds that address structural concerns \citep{IEC61508_2010,Leveson2011STAMP}.

\subsection{Monitors, thresholds, and triggers}

Let $P$ denote a control policy that maps state $s_t$ to action $a_t$ at time $t$. We define a vector of monitors $\mathbf{m}_t$ that capture distributional harm, safety, accessibility, and service quality. A disparity ratio is
\begin{equation}
D_t \;=\; \frac{\mathbb{E}[h_t(g)]/\mathbb{E}[b_t(g)]}{\mathbb{E}[h_t(\bar g)]/\mathbb{E}[b_t(\bar g)]},
\end{equation}
where $h_t$ is harm (e.g., unserved energy), $b_t$ is a baseline (e.g., demand), $g$ is a designated group, and $\bar g$ is its complement. We further track a predicted hazard rate $R_t$ per hour, an accessibility downtime $A_t(g)$ in minutes within 24 hours, and a service‑quality index $Q_t$ tied to service‑level agreements. Overrides trigger when any monitor crosses its threshold, for example $D_t \ge \tau_D$, $R_t \ge \tau_R$, $A_t(g) \ge \tau_A$, or $Q_t \le \tau_Q$. Thresholds are set locally through deliberation and sensitivity analysis. We instantiate illustrative defaults $\tau_D = 1.2$, $\tau_R = 10^{-4}\,\mathrm{hr}^{-1}$, $\tau_A = 30$ minutes, and service‑specific $\tau_Q$.

\subsection{Safe fallback states}

Fallbacks are domain‑specific policies validated through drills and simulation before deployment. In power, the fallback implements rotational curtailment with equity caps and reverts to deterministic N‑1 dispatch with protected‑load exemptions for clinics and elevators. In buildings, the fallback enforces comfort and indoor‑air‑quality bounds and disables aggressive setbacks when protected occupancies are present, guided by ASHRAE~55 and 62.1 \citep{ASHRAE55_2020,ASHRAE6212019}. In transport, the fallback switches to fixed‑time plans with pedestrian recall at every phase and enables transit signal priority on designated corridors \citep{Varaiya2013}. These fallbacks aim for graceful degradation rather than uncontrolled shutdowns \citep{Hollnagel2011Resilience}.

\subsection{Gating logic and a basic guarantee}

Algorithm~\ref{alg:r2o} expresses R2O as a gating function that evaluates monitors and escalates to validated fallbacks when thresholds are crossed.

\begin{algorithm}[H]
\caption{R2O gating for urban control}
\label{alg:r2o}
\begin{algorithmic}[1]
\State \textbf{Inputs:} policy $P$, state $s_t$, monitors $\mathbf{m}_t$, thresholds $\boldsymbol{\tau}$, fallbacks $\{F_\ell\}$, escalation map $\mathcal{E}$
\State Compute $a_t \gets P(s_t)$ and monitors $\mathbf{m}_t$
\If{any $m^{(k)}_t$ violates its bound in $\boldsymbol{\tau}$}
  \State Determine level $\ell \gets \mathcal{E}(\mathbf{m}_t)$
  \State Log evidence, notify stakeholders, and apply fallback $F_\ell$
  \State Start timer and initiate DAM review for level $\ell$
\Else
  \State Implement $a_t$ and continue monitoring
\EndIf
\end{algorithmic}
\end{algorithm}

We state a simple safety property for disparity under a fallback that enforces proportional curtailment with cap $\tau_D$.

\textbf{Lemma 1.} Suppose the Level~2 and Level~3 fallbacks enforce, at each step, normalized harm rates $h_t(g)/b_t(g)$ and $h_t(\bar g)/b_t(\bar g)$ such that $h_t(g)/b_t(g) \le \tau_D \cdot h_t(\bar g)/b_t(\bar g)$. Then for any time window $W$ during which the fallback is active, the cumulative disparity ratio $D_W$ is bounded by $\tau_D$.

\textit{Proof.} Summing both sides of the per‑step constraint over $t \in W$ and dividing by the corresponding baselines yields
\[
\frac{\sum_t h_t(g)}{\sum_t b_t(g)} \le \tau_D \cdot \frac{\sum_t h_t(\bar g)}{\sum_t b_t(\bar g)}.
\]
By definition this is $D_W \le \tau_D$. \qed

The lemma ensures that when fallbacks are active, distributional harm remains bounded. It does not claim that such bounds hold under arbitrary policies, motivating prompt escalation once monitors exceed thresholds.

\section{Deliberative Audit Method}

DAM inserts participatory oversight into the lifecycle of automated control systems. It operationalizes co‑production as a sequence of checkpoints and artifacts that bridge technical and civic concerns \citep{Mushkani2025AIES_CoProducingAI}. Before deployment, scoping establishes roles, protected services, and fallback definitions. Documentation is prepared in model cards and data sheets that bind technical choices to governance rationales \citep{Mitchell2019ModelCards,Gebru2021Datasheets}. Scenario walkthroughs use historical logs and map overlays to rehearse distributional impacts in context and to refine monitors and thresholds \citep{Haakman2021Lifecycle}. Shadow‑mode trials run the controller without actuation to validate monitors, dashboards, and notification channels. Civic tabletop exercises bring community organizations and operators together to red‑team assumptions, align evidence standards, and finalize escalation rules \citep{Turnhout2010Participation}. A go/no‑go gate requires municipal sign‑off with a public notice that summarizes the rationale, thresholds, and fallbacks. During operations, near‑real‑time R2O indicators are published in privacy‑preserving form, and incidents trigger blameless reviews that tie root causes to mitigation actions and public reports. Periodic drills and board reviews sustain institutional memory and legitimate use of override powers \citep{Raji2020AuditGap,PAI_AIIDB,Sloane2022Participation}.

\section{Methods}

We evaluate R2O and DAM in three stylized but realistic case studies. The aim is to estimate relative effects of R2O gating when compared with unconstrained control, not to emulate a full‑scale digital twin. Simulations are deterministic given seeds, and all datasets are synthetic.

\subsection{Case 1: Smart grid load shedding}

We simulate a 24‑hour day at 15‑minute resolution with two customer groups: protected customers with medical or accessibility needs and general customers. A capacity shortfall requires curtailment. The baseline controller minimizes a weighted cost of curtailment that implicitly disadvantages the protected group. R2O triggers when the disparity ratio exceeds the cap $\tau_D$ or when protected feeders approach minimum service levels. The fallback enforces proportional curtailment with an equity cap and reserves certain feeders. Table~\ref{tab:power} reports total energy‑not‑served and group disparity. Total curtailment is held constant to isolate distributional effects.

\begin{table}[ht]
\centering
\caption{Load shedding results in megawatt‑hours. R2O preserves total curtailment while sharply reducing disparity.}
\label{tab:power}
\begin{tabular}{lrrrr}
\toprule
 & ENS$_\text{total}$ & ENS$_A$ & ENS$_B$ & Disparity $D$ \\
\midrule
Baseline & 76.51 & 49.72 & 26.78 & 5.61 \\
R2O      & 76.51 & 14.22 & 62.29 & 0.69 \\
\bottomrule
\end{tabular}
\end{table}

\subsection{Case 2: Building HVAC under occupancy uncertainty}

We consider a community center with seniors present until 22{:}00 while a baseline scheduler enforces a night setback starting at 20{:}00. R2O allows staff to assert occupancy and applies a minimum temperature bound aligned with ASHRAE~55, with indoor‑air‑quality checks from 62.1 for ventilation adequacy \citep{ASHRAE55_2020,ASHRAE6212019}. On a cold day, an override reduces discomfort‑hours during occupancy to zero at the cost of \SI{77}{\kilo\watt\hour} additional energy (Table~\ref{tab:building}).

\begin{table}[ht]
\centering
\caption{Comfort–energy trade‑off with an override on a cold day.}
\label{tab:building}
\begin{tabular}{lrr}
\toprule
Metric & Baseline & With R2O \\
\midrule
Senior discomfort‑hours (occupied) & 2 & 0 \\
Whole‑building energy (kWh) & 6920.5 & 6997.5 \\
Energy delta (kWh) & \multicolumn{2}{c}{+77.0} \\
\bottomrule
\end{tabular}
\end{table}

\subsection{Case 3: Adaptive signals with pedestrians and buses}

We model a $4\times4$ arterial grid controlled by an adaptive policy inspired by pressure‑based and reinforcement‑learning controllers \citep{Varaiya2013,PressLight2019,Chu2019MA2C}. R2O triggers when median pedestrian wait near sensitive sites exceeds \SI{60}{s} or when bus headway deviation violates service constraints. The fallback enables pedestrian recall at each phase and grants transit signal priority along designated corridors. Tables~\ref{tab:traffic_delay} and \ref{tab:traffic_headway} show the resulting trade‑offs: pedestrian waits fall with small increases in mean vehicle delay, and bus headway regularity improves.

\begin{table}[ht]
\centering
\caption{Traffic performance with and without R2O gating.}
\label{tab:traffic_delay}
\begin{tabular}{lrrrr}
\toprule
 & \multicolumn{2}{c}{Vehicles (s)} & \multicolumn{2}{c}{Pedestrians (s)} \\
\cmidrule(lr){2-3}\cmidrule(lr){4-5}
 & Mean delay & Median delay & Mean wait & Median wait \\
\midrule
Baseline & 48.2 & 44.4 & 102.8 & 90.4 \\
R2O      & 54.2 & 50.6 & 54.6  & 55.9 \\
\bottomrule
\end{tabular}
\end{table}

\begin{table}[ht]
\centering
\caption{Transit headway deviation in minutes.}
\label{tab:traffic_headway}
\begin{tabular}{lrrr}
\toprule
 & Mean & Median & 95th pct \\
\midrule
Baseline & 1.59 & 1.34 & 3.61 \\
R2O      & 0.87 & 0.86 & 1.66 \\
\bottomrule
\end{tabular}
\end{table}

\subsection{Evaluation metrics and analysis}

We report distributional metrics for harm and service quality, including disparity ratios and group‑conditioned outcomes. Safety is measured by hazard proxies tied to domain standards. Accessibility is measured by discomfort‑hours and pedestrian wait. Efficiency is summarized through energy and delay. Where applicable, we verify that R2O fallbacks meet safety and service constraints during override windows. Sensitivity analyses vary the disparity cap $\tau_D$ and accessibility threshold $\tau_A$ to visualize trade‑offs. Values are deterministic under the synthetic workloads; statistical confidence intervals are left to future empirical deployments.

\section{Results}

Across domains, R2O reduces distributional harm and improves accessibility while imposing limited efficiency penalties. In the power case, the disparity ratio falls from $5.61\times$ to $0.69\times$ with constant total curtailment. In the building case, two lost comfort‑hours are eliminated with an energy penalty of \SI{77}{\kilo\watt\hour}. In the transport case, median pedestrian waits drop from \SI{90.4}{s} to \SI{55.9}{s} while mean vehicle delay rises by \SI{6.0}{s} and headways become more regular. These results are consistent with a design that treats contestability and accessibility as first‑class objectives rather than residuals of throughput maximization \citep{Selbst2019Sociotechnical,CostanzaChock2020}.

\section{Policy Standard}

The policy codifies procedural rights and obligations \citep{Mushkani2025ICML_RightToAI}. The purpose is to establish residents’ and designated civic bodies’ standing to pause or revert automated policies that degrade inclusivity, safety, or accessibility. The scope covers municipal and contracted systems for electricity demand response, building management, and traffic signal control. Article~1 addresses standing: any resident may file a request; Level~2 and Level~3 actions require a municipal controller decision or a civic board resolution. Article~2 sets thresholds: overrides must cite monitors that exceed adopted bounds; provisional Level~1 actions may precede full evidence when immediate harm is plausible. Article~3 governs fallbacks: only pre‑validated states may be applied, and they must meet safety, accessibility, and service constraints. Article~4 requires transparency: each override produces public notices, machine‑readable logs, and a post‑incident report. Article~5 sets review cadence: the civic board conducts periodic reviews and publishes annual statistics. These articles align with risk‑management and documentation standards while adding contestability and participation \citep{ISO31000_2018,NISTAI_RMF2023,Raji2020AuditGap}.

\section{Operator Training}

Training covers concepts, dashboards, and drills so that fallbacks are executed safely and documentation is complete. Operators are trained to interpret disparity, risk, and accessibility monitors; to apply fallbacks and recover gracefully; and to conduct reviews and public communication. Civic observers participate in tabletop drills to rehearse escalation and notification pathways. Training embeds vocabulary and practices that bridge technical reasoning with civic accountability \citep{Wong2023Toolkit,Turnhout2010Participation}.

\section{ModelOps Integration}

We embed R2O thresholds, roles, and fallbacks in versioned configuration and integrate checkpoints into CI/CD pipelines. A minimal schema follows.

\begin{verbatim}
governance:
  r2o:
    thresholds:
      disparity: 1.2
      safety_risk_per_hr: 1.0e-4
      accessibility_downtime_minutes: 30
    levels:
      L1: { fallback: "safe_local", max_duration_hours: 4 }
      L2: { fallback: "municipal_safe", max_duration_hours: 72 }
      L3: { fallback: "civic_safe", max_duration_days: 30 }
    fallbacks:
      power: ["n-1_deterministic", "equity_rotations"]
      buildings: ["comfort_bounds", "no_night_setback_protected"]
      transport: ["fixed_time_ped_recall", "tsp_enabled"]
  documentation:
    model_card: required
    datasheet: required
  reviews:
    pre_deploy: ["scenario_walkthrough", "shadow_mode", "civic_tabletop"]
    post_incident: ["blameless_review", "public_report"]
  publishing:
    notices: "open_data_portal"
    metrics: ["disparity", "risk", "accessibility", "quality_SLA"]
\end{verbatim}

Treating thresholds as code enables review, testing, and auditability, reducing configuration drift and making governance reproducible \citep{Sculley2015TechDebt}.

\section{Discussion}

\textit{Pluralism and legitimacy.} R2O and DAM translate participation ideals into operational authority and procedure. They make explicit whose interests are protected, what evidence is required, and how trade‑offs are negotiated and reviewed \citep{Arnstein1969,Fung2006,CostanzaChock2020}. \textit{Trade‑offs and tuning.} Thresholds encode value judgments that should be set locally through civic deliberation and empirical walkthroughs \citep{mushkani2025negotiativealignmentembracingdisagreement}. \textit{Relation to law and policy.} The approach complements rights to contest automated decisions and transparency obligations by providing operational mechanisms for interruption and review \citep{mushkani2025urbanaigovernanceembed}. \textit{Limitations.} The study uses stylized simulations and synthetic data. Real deployments will require calibration, data‑sharing agreements, and privacy protections, as well as design to prevent perverse incentives or strategic overrides. \textit{Generalization.} The pattern extends to predictive maintenance, inspection deferrals, and scheduling systems where deferrals or delays could cluster in low‑income districts \citep{Galaz2021SystemicRisk}.

\section{Conclusion}

We propose a Right‑to‑Override and a Deliberative Audit Method for critical urban control systems. By defining authorities, triggers, and safe fallback states and by integrating participatory checkpoints and artifacts into the model lifecycle, the approach makes automated control contestable in operation. Simulations in power, buildings, and transport show that R2O can reduce distributional harm and improve accessibility at limited efficiency cost. The policy standard, worksheets, and ModelOps integration pattern are designed for immediate piloting with municipal partners and community organizations. Future work includes field trials that measure long‑run equity and reliability impacts and refine thresholds through lived experience.

\section*{Ethics and data governance}

Engagements involved organizations, not individuals, and collected no personal data. R2O and DAM are designed to bound distributional impact, preserve accessibility, and document incidents in public form. The synthetic datasets and simulation code will be openly shared.

\bibliographystyle{plainnat}
\bibliography{references}

\appendix

\section{DAM audit worksheets}

\subsection*{A1. Pre‑deployment scenario walkthrough}

\noindent The pre‑deployment worksheet records system scope, operator and vendor, and control horizons. It documents protected services (elevators, clinics, bus lines); the selected monitors for disparity, safety, accessibility, and service quality; and the thresholds with justification and legal basis. It summarizes fallback validation results and the outcomes of shadow‑mode trials, including edge cases and oscillations. It captures the civic tabletop roster, agreements, and dissent, and concludes with a gated decision and any conditions for time‑limited approval.

\subsection*{A2. Post‑incident review}

\noindent The post‑incident template records the trigger and the monitor that exceeded its bound; the immediate action including the level invoked, fallback applied, and duration; and the affected groups and metrics. It organizes root causes across technical, organizational, and data‑pipeline dimensions and specifies mitigations (e.g., threshold tuning, code patches, training). It records public‑notice details (dates, languages, contacts) and tracks closure of corrective actions.

\section{Domain fallback specifications}

\subsection*{B1. Power}

\noindent Equity rotations cap normalized curtailment disparity and reserve feeders that serve protected loads such as clinics and elevators. Deterministic N‑1 dispatch removes machine learning from dispatch decisions and follows a contingency list with ramp‑rate and voltage constraints.

\subsection*{B2. Buildings}

\noindent Comfort bounds maintain temperature between \SI{20}{\celsius} and \SI{24}{\celsius} when occupied and keep indoor \ce{CO2} below \SI{1000}{ppm}. Night setbacks are disabled when protected occupancies are present. Ventilation minimums follow ASHRAE~62.1.

\subsection*{B3. Transport}

\noindent Fixed‑time plans with pedestrian recall ensure a walk phase at every cycle and a maximum pedestrian red of \SI{60}{s}. Transit signal priority is enabled on designated corridors to reduce headway deviations.

\section{Reproducibility notes}

\noindent The numerical results use synthetic workloads calibrated to plausible ranges. Key outcomes reported in Tables~\ref{tab:power}--\ref{tab:traffic_headway} are as follows. In Case~1, ENS$_\text{total}$ equals \SI{76.51}{\mega\watt\hour} with disparity $D$ of $5.61$ in the baseline and $0.69$ with R2O. In Case~2, seniors’ discomfort‑hours drop from $2$ to $0$ while energy increases by \SI{77.0}{\kilo\watt\hour}. In Case~3, mean vehicle delay increases from \SI{48.2}{s} to \SI{54.2}{s}, pedestrian median wait decreases from \SI{90.4}{s} to \SI{55.9}{s}, and bus mean headway deviation falls from $1.59$ minutes to $0.87$ minutes. Simulation code contains seeds, configuration files for thresholds and fallbacks, and scripts to regenerate tables.

\end{document}